# Practices and challenges in clinical data sharing

Fida K. Dankar

Computer Science department, NYU Abu Dhabi

fd2242@nyu.edu

Abstract. The debate on data access and privacy is an ongoing one. It is kept alive by the never-ending changes/upgrades in (i) the shape of the data collected (in terms of size, diversity, sensitivity and quality), (ii) the laws governing data sharing, (iii) the amount of free public data available on individuals (social media, blogs, population-based databases, etc.), as well as (iv) the available privacy enhancing technologies. This paper identifies current directions, challenges and best practices in constructing a clinical data-sharing framework for research purposes. Specifically, we create a taxonomy for the framework, identify the design choices available within each taxon, and demonstrate thew choices using current legal frameworks. The purpose is to devise *best practices* for the implementation of an effective, safe and transparent research access framework.

## 1. Introduction

Advances in health information technology have enabled the collection and storage of large amounts of clinical data. Access to this data for research is restrained, although vital for the development of evidence-based and personalized healthcare. Recently, some governmental institutions were enacted with a (dedicated) mandatory health data-sharing law for research purposes [1,2].

Globally, health data platforms are becoming essential infrastructures in our societies. They are responsible for releasing data with third parties for research purposes in an ethical and lawful manner. This entails the application of acceptable *privacy-enhancing processes* to regulate the risk of privacy loss as well as the implementation of proper *data-access control and oversight* mechanisms that fulfill applicable legal requirements.

De-identification is the main privacy-enhancing technology used in health data-sharing to control privacy risks [3,4]. It designates the process of removing identifying information from the data, thereby removing the association between the data and the subjects. However, there are growing concerns that the rise in publicly available information online has made it possible to re-identify what was thought to be properly de-identified data [4–6]. Moreover, the increasing adoption of medical health records and its aggregation with data from multiple other sources (such as genetic data, environmental data and clinical trial data) created more challenges to de-identification [4,6,7]. This, in addition to the adoption of stricter legislations with respect to data sharing in the absence of subjects' consent [8,9]. Thus, the problem of designing a proper privacy mechanism for the usage of clinical data (although not new) is still a challenging problem.

In this manuscript, we identify the *key challenges* in the usage of clinical data for research purposes in the context of a central health data repository. The purpose is to devise *best practices* for the implementation of an effective research access framework. The contributions of the paper are the following:

- First, we distinguish 5 principal (interrelated) modules/taxon in the design of a research access framework. Each module is related to one aspect of the framework design: governance structure, research database set-up, access control mechanism, access requests endorsed and user support.
- We identify the key design decisions within each of the identified modules, whereby different decisions result in diverse governance models.
- We present 4 examples of current well-established research access frameworks offering various design choices.
- Lastly, we formulate best practices and recommendations for an effective research access framework.

The manuscript is divided as follows: Section 2 defines the concepts of re-identification risk and data identifiability; Section 3 presents a taxonomy of the framework design and presents the available choices within each taxon based on current practices in the field. Section 4 examines four current frameworks from the field. We conclude the paper in Section 5 with a summary of the results and recommendations on best practices.

## 2. Methods

### 2.1. Background

Re-identification is performed by linking de-identified dataset with publicly available information using attributes that are common to both (as displayed in Figure 1) [5,10,11]. Many public datasets have been used for re-identification purposes, examples include the Internet Movie Database IMDB [12], mobility data from internet/phone usage [13,14], and publicly available genome datasets [6].

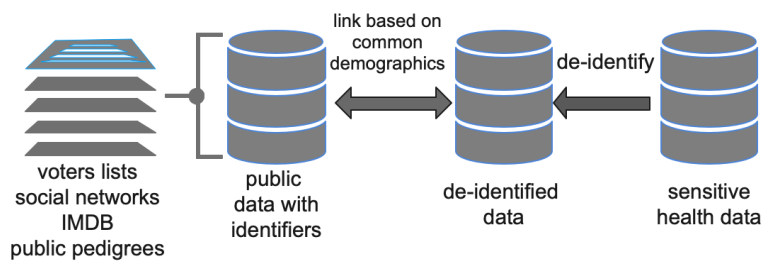

Figure 1. Re-identification process.

The problem is becoming more prevalent due to the significant growth in the amount of public data available on the internet to link with, as well as the amount and specificity of clinical data collected, thereby, providing more material for a successful linkage attack. On the other hand, it is becoming harder to evaluate the re-identification risk (or identifiability) of datasets as we cannot foresee the amount/type of data that attackers may have/use in their de-identification attempts.

#### 2.1.1. Identifiability spectrum

The Canadian Anonymization Network (CANON) defined recently a spectrum of identifiability for information [15]. It consists of 3 states with hazy boundaries: identified information, Identifiable information and non-identifiable information. Identified information is information that directly identifies an individual, identifiable information is information with a serious possibility of re-identification for one or more individuals, and non-identifiable information is information with no serious possibility of re-identifying (one or more) individuals [15].

In [16], Gellman introduced another category of information- potentially identifying information (PI)- which he defined as information that is potentially identifiable.

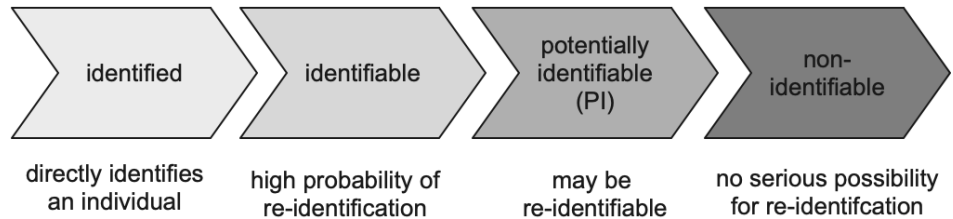

Figure 2. State of information.

According to Gellman, and for the same aforementioned reasons, it is not easy to determine whether information is re-identifiable at any point in time, thus, all information with non-zero utility that is not obviously identifiable, should be considered as PI. Accordingly, zero re-identification risk can never be guaranteed [17]. Alternatively, one can attempt to lower the risk through understanding of the factors that affect re-identification [18]. Figure 2 summarizes the suggested states (note that utility is proportional to identifiability).

### 2.1.2. Re-identification risk

Re-identification risk depends on multiple factors that are difficult to quantify, in addition to (the aforementioned) inability to estimate the data available to the attacker, the risk is dependent on the de-identification mechanism used, the willingness and capacity of individuals to mount a re-identification attacks as well as the sensitivity of the information within the de-identified dataset [18,19] (willingness to re-identify maybe higher with more sensitive data).

Re-identification risk was formally defined in [20] as the probability of attempting a re-identification and that such attempt is successful (Figure 3). Mathematically:

$$Prob(re-identification) = Prob(re-identification/attempt) * Prob(attempt) \quad 3.1$$

Thus, controlling risk amounts to either lowering the probability of an attempt or its probability of success [19]. The probability of success, referred to as *data risk*, is a function of the de-identification applied on the dataset as well as the public available for linkage. While the risk of waging a re-identification attack, referred to as *context risk*, depends on contextual factors such as the benefit that would be gained from such re-identification, the difficulty associated with bypassing installed security and the penalty applied to breaking any associated data sharing contracts [21,22].

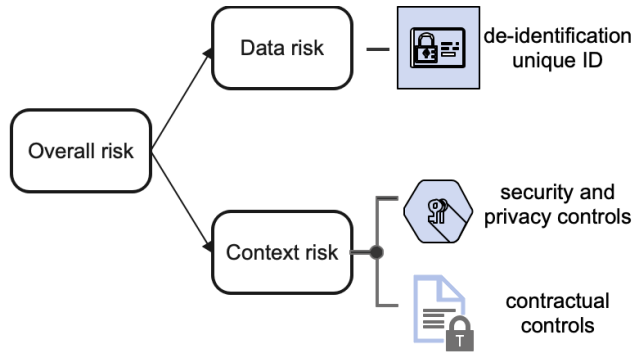

Figure 3. Re-identification risk representation. Data risk represents the Prob(re-identification/attempt) and the context risk represents the probability of an attempt

With the above formulation, controlling re-identification risk entails the control of both data risk and contextual risk. Data risk can be controlled via one of the recognized de-identification and/or data protection mechanisms taking into consideration the particular environment at hand (laws, public data availability, and population distributions). Context risk can be controlled through well designed governance structures, access controls, and data use agreements among others. As such, the implementation of a privacy-compliant data sharing framework is a challenging multi-decision undertaking. The different key decisions and their impact will be discussed in details in the remainder of the manuscript. We start first by presenting the design framework taxonomy. The taxon are not independent of each other, in that decisions are interrelated and may have to be done simultaneously, however, it makes for a simpler and more organized presentation of the different ideas.

## 2.2. Design elements taxonomy

As mentioned before, implementing a data sharing framework is a challenging task. It requires choices with respect to multiple interrelated issues. For many researchers, the first step in this process is to study the applicable laws and policies and to examine the processes of similar projects in the field [23]. In fact, several warehouses have current and well-established structures for research-based clinical data sharing. They differ in multiple key aspects: (i) the data flow and governance structure, (ii) the clinical research database set-up, (iii) the data access model, (iv) the user requests endorsed, and (v) the user support offered.

We refer to this categorization as the framework taxonomy, and we identify the key decisions related to its 5 taxon/aspects, then present the practices and challenges related to their implementation before providing real use cases from a diversity of clinical data sharing models. We limit our context to one central data repository where data is linked and stored with a purpose limited to scientific research.

The five taxons along with their associated decisions are depicted in Figure 4 and discussed separately in the subsequent sections.

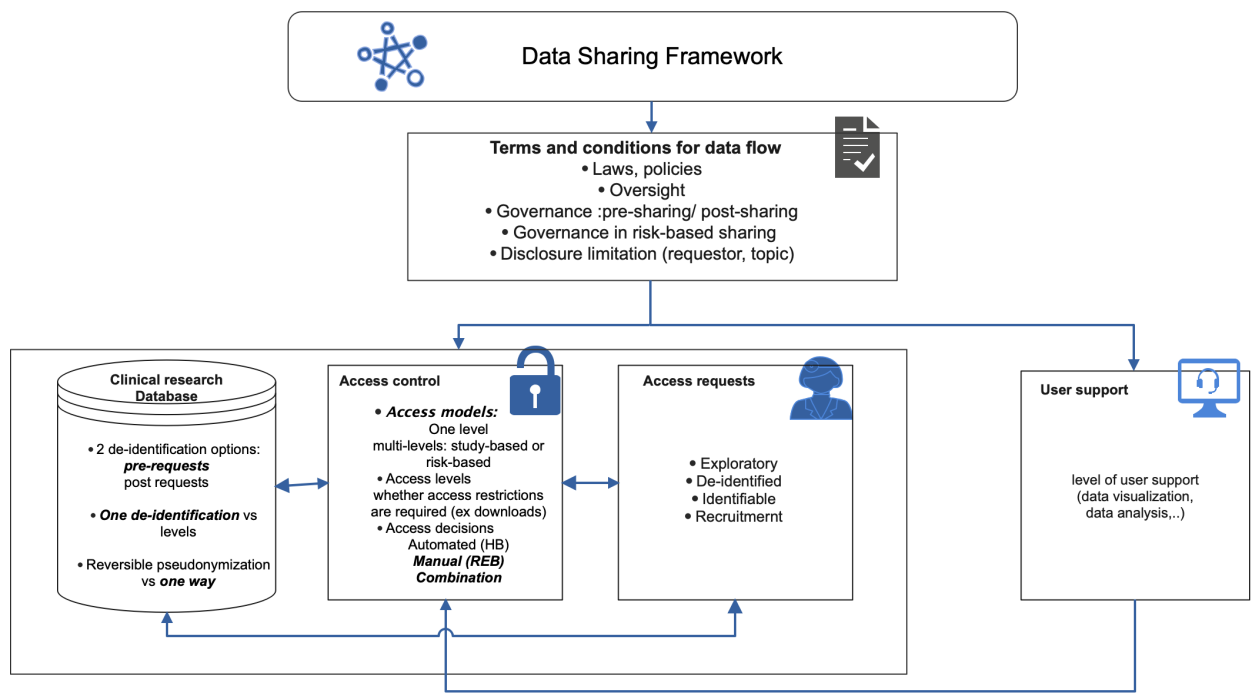

Figure 4. Taxonomy of the research framework. Arrows indicate dependency between the taxon. taxon inside the box are highly interdependent.

### 2.2.1. Data flow and governance structure

One of the first tasks in a data sharing framework implementation is the identification and implementation of a governance structure. Data Governance is the backbone of the access framework, it is the set of processes, and procedures implemented by the health data holder to manage and share their data while guaranteeing its protection across all development and usage stages. Data governance lays down mechanisms for (i) oversight structures and responsibilities, (ii) data collection, protection and update, (iii) processes and policies for data access, (iv) public engagement and process transparency and (v) user support. Setting a governance structure requires the engagement of ethicists, and lawyers to make sure that decisions made adhere with applicable laws, policies, and best practices. The definitions of the different issues along with the associated decisions are explained in *Table 1*.

Table 1. Some of the key decisions associated with the design of a governance structure

| Category | Associated decisions | Best practices |
|---|---|---|
| *Oversight, responsibilities and accountability* (Defines the different layers of oversight that ensure accountability (at all levels) and trust | *Process oversight:* How many layers of oversight? Will the public be part of the oversight process? How independent are the different oversight committees? | Tiered oversight composed of independent institutions. Community involvement in oversight. Enforcing penalties in case of noncompliance (with proportionate severity). |
| in the overall framework.) | What penalties are applied in case of non-compliance? *Customers oversight:* How to ensure users compliance with the terms of data use? Should the oversight be proportionate to the risk induced by the data sharing episode? | Customer surveillance post data sharing (proportional to data sensitivity). Periodic evaluation of oversight mechanisms (to include new findings/technology) References: [4,24–26]. |
| *Data collection, Data protection, data updates and reporting* | Are health institution required by law to share their data? What are the methods to employ for data collection and aggregation? What is the scope of consent? Data preprocessing methods (including pseudonymization) What data protection regime will be used? Will it be adaptive (request dependent) or fixed? And if fixed, do we have one fixed de-identified dataset or multiple datasets with different de-identification levels? What level of identifiability are users allowed to request? (de-identified, identifiable, aggregate data, synthetic,...) | Standardized methods for data generation that ensure data quality Best practices in data security are implemented to reduce breaches Pseudonyms (unique patient IDs) are necessary for data updates. They are made reversible in case there is a need to inform specific participants with relevant information about their condition or to report incidental findings. References: [26–28]. |
| *Processes and policies for data access* | What are the disclosure rules? What is the approval process? (how many committees should a user seek approvals from? Is this process fixed or does it depend on the particular request/user?) Will the de-identification be dependent on the risks involved by the data request? How will data usage be designed? (fee for service, not for profit,…) | Approval process should be fair and transparent Fixed (preset) time limits on approval process duration. Comprehensive request-based adaptive data access regime, where requests are dealt with on a case by case basis. References: [1,26,28,29]. |
| *Public communication, and process transparency (including return of actionable results to patients)* | What public engagement process will be followed? How involved will the community be in the design and delivery of the framework? Will any results be returned to users? | Public involvement in every step of the process Clear and well-defined data access rule Data publicly accessible according to the pre-set rules |

| | | Periodic updates on general research achievements and future goals |
| | | References: [26,30–32] |
| *Users' / community support* | How will users/community be encouraged to use or be involved in the framework? What kind of support will be provided to the users? | Awareness within the community should be a priority and engagement facilitated. Researchers should have the tools and resources necessary to conduct their research. References: [4,25,33] |

### 2.2.2. Clinical research database set-up

Data curators have the option to maintain a de-identified dataset and use it to serve all data request, or they can tailor de-identification in response to individual requests received. The first option leads to faster response rate and higher efficiency, while the second provides better quality as the de-identification may be adapted to the needs of the investigator.

To benefit from both, efficiency and quality, some frameworks maintain multiple de-identified datasets with different forms/levels of de-identifications. However, this increases security and data management burdens.

The de-identification process often starts with the creation of a research unique identifier for each subject, RUI (also known as pseudonym), and the removal of all directly identifying information such as name, IDs and emails (as defined in the HIPAA limited dataset [3]). The RUI is often obtained by applying a secure cryptographic function to one of the unique patient identifiers, such as medical record number or citizenship number [4]. When no unique identifier is available, a combination of identifiers maybe used to create the RUI (for example name along with date of birth) [34].

RUIs are used for the purpose of updating de-identified records with new data when it becomes available. The cryptographic RUI function can be reversible or non-reversible. Reversible functions offer the possibility of reversing the de-identification and thus contacting subjects to request additional data or to inform them of actionable incidental findings, or breakthroughs relevant to their health [27].

### 2.2.3. The data access model

Data sharing mechanisms can be categorized into two broad categories: open-access and controlled-access. Controlled-access models operate on a data request basis and impose necessary controls/safeguards on data access [18]. Open-access models often operate under a mandate from participants (who want to publish their data to advance science) or under the assumption that the data is anonymized. The definition of anonymization differs among privacy laws, it is often used interchangeably with de-identification. Recital 26 of GDPR defines anonymous data as "information which does not relate to an identified or identifiable natural person, or to personal data rendered anonymous in such a manner that the data subject is not or no longer identifiable". In other words, one needs to consider the realistic means that may be used to re-identify a person (such as cost, time required and technology) when assessing anonymization [35]. In this paper, we categorize de-identification as an attempt for achieving anonymization.

Because of the highly sensitive nature of health data, and due to the unceasing failures of de-identification attempts [2,36], access regimes generally employ controlled-access. They control data risk and context risk while adhering to applicable laws. These laws tend to specify the (minimal) necessary conditions and safeguards under which health data sharing can operate. Multiple controlled access models have been developed worldwide. We present some of these models below and provide some implementations in section 3 spanning multiple countries/continents.

*One size fits all:* A simple approach for data sharing is the One size fits all approach. The research database is de-identified and an Institutional Review Board, IRB, also known as ethical review board, reviews all data sharing applications. Only when it is determined that the benefits of the study outweigh the risks can the request be approved (Figure 5).

Platforms implementing this approach suffer several shortcomings. They are not flexible, they cannot embed all the legal and ethical regulations in the biomedical domain, and, in particular, they do not recognize that the de-identification risk and potential of harm vary across user requests (for example: a request for covid 19 data vs a request for mental health data) and thus are unable to impose stricter constraints on data accesses that exhibit higher privacy risk in terms of additional de-identification or access controls (which results in denying such requests).

Current frameworks are moving away from this approach by extending their models to include multiple de-identified databases [25,31,37].

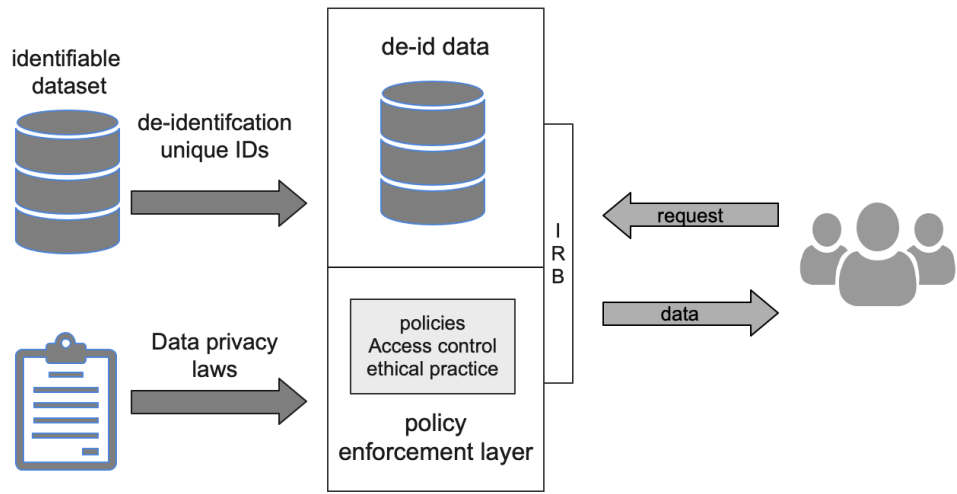

Figure 5.One size fits all model. De-id refers to de-identification.

*Risk based access:* Risk-aware access control attempts to quantify the risk posed by a data request and design models that impose more mitigation measures on the requests with higher risk. Such mitigations could manifest as reduction in the specificity of the data (reducing data risk) and/or as restrictions on users access to the data (reducing context risk). Thus, a risk-based data sharing model requires modeling the privacy risk posed by a data request, specifying multiple data protection levels, and establishing decision rules for mapping computed risks to appropriate protection levels [18]. Multiple efforts were targeted at quantifying data privacy. In [38], the authors introduced a 4 dimensional model for assessing the privacy risk of a data request: the purpose of the request (data uses), the visibility (who will access the data), the granularity (data specificity) and the retention (total time the data will be kept in storage). In multiple consecutive studies [19,39], the model was updated to include users' motives (user risk), and sensitivity of the requested data (data risk). In [18], the authors extended the model further to include institute risk (security controls employed at the institution to which the user is affiliated as well as its reputation). The authors also provided multiple data protection levels by varying the de-identification parameters (for example, different values of $k$ and $\epsilon$ provide different anonymity levels in k-anonymity and differential privacy respectively [40,41]), and varying data release levels as follows(refer to Figure 6(a)):

1. The release and forget level: where data is released to the user and control over the data is lost
2. Data use agreement level: the data is made available on specific terms that specify purpose and legal usage
3. Remote access level: de-identified record-level data is kept in a secure platform and accessed remotely. User are able to run analysis remotely without the ability to download the data.

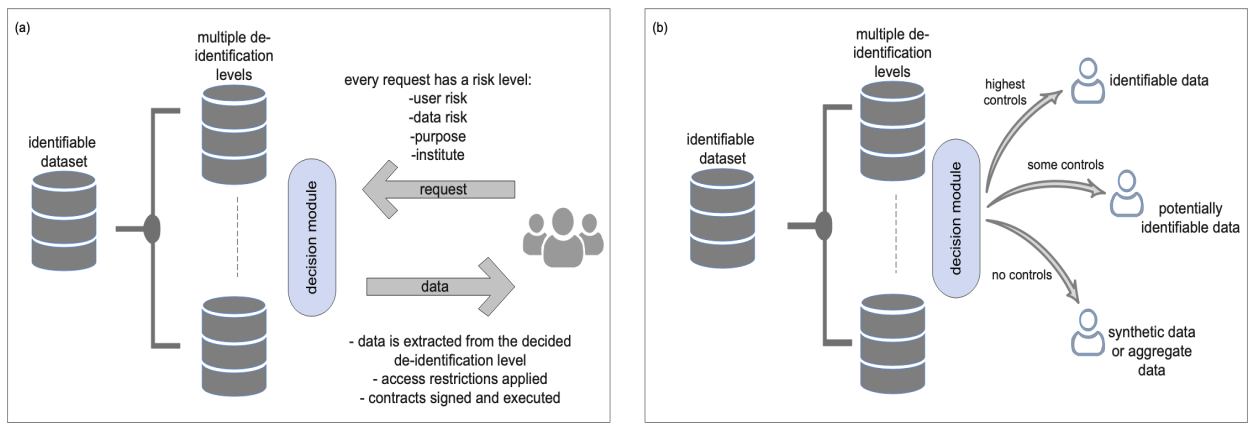

Figure 6. (a) Risk-based model (b) risk-based model where risk is defined in terms of data identifiability

Quantification of data risk is a difficult task to achieve. In fact, calculating the different risk-dimensions requires a mechanism for recording the users/institutions reputation (such as certifications and histories). Moreover, it requires the calculation of the sensitivity of the requested data which could be subjective and difficult to measure. As such, common variations of the model base their decisions on one of the risk dimensions, such as (i) the institution to which the user is affiliated (for example is it public or private?), or (ii) the granularity of the data requested by the user (for example is it aggregated data, de-identified data or identifiable data?). Figure $6$(b) depicts a scenario where risk is proportional to the granularity of the requested data. Different risk categories are met with mitigation measures/controls to counter the posed risk. As synthetic and aggregate data present lower risk than PI data, they are met with the minimal controls [42].

*Adaptive:* Adaptive data sharing models employ post-request de-identification. An identifiable or lightly de-identified database is maintained, and an appropriately de-identified dataset is created for each approved data request on the basis of data minimization. Such process could require negotiation between the data curators and users to define and agree on the best data for the study, refer to Figure 7.

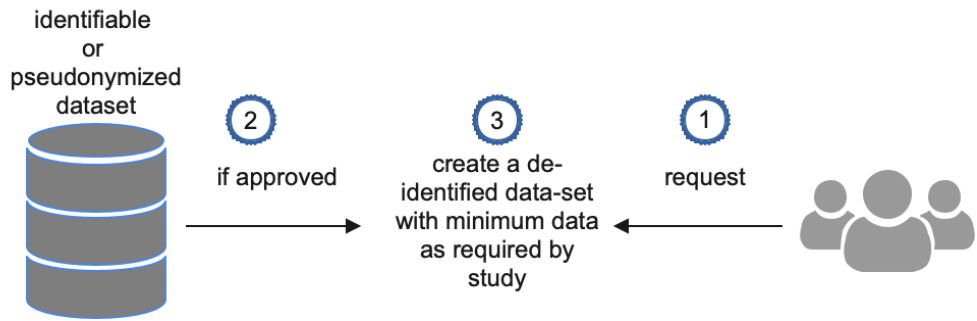

Figure 7. Adaptive models

In all aforementioned models, review boards are required to study individual data-access applications in order to assess their risk. Such evaluation causes delays for research activities resulting in sustainability problems over the long run. Possible solutions include IRB automation [43,44], and fees for service models [45]. The Risk-based models offer a good platform for automation as argued in multiple investigations [18,39,46–48].

### 2.2.4. Type of requests endorsed

The type of user requests to support is another decision to be tackled. A choice of whether identifiable data will be offered for certain studies or whether data sharing will be restricted solely to de-identified data is an essential one. For cases where identifiable data is to be shared, the conditions, oversight and access restrictions on such data needs to be well defined and subjects made aware of the involved risks. Another form of requests to manage is the cohort identification requests. In some implementations, such requests require approvals from ethics boards (even when only count queries are sought) [49], Other implementations allow obfuscated to queries and limit their numbers per user to thwart re-identifications [25].

### 2.2.5. User support offered

Research on modern clinical and biomedical data is highly complex and requires diverse range of expertise. Offering the highest level of support at diverse phases of the research project life cycle would enrich and involve the warehouse scientist with the variety of research projects. However, it adds the burden of maintaining a team with deep and diverse research and clinical expertise. Some warehouses have opted to employ extensive user support units [25], some provide the necessary software for the users to develop their own analysis tools [33], while others provide no support (mostly in frameworks that release the data to the users, or during early phases of development) [37,50,51].

## 3. Results- use cases

This section present examples of established health data sharing frameworks. The selection spans four countries: UK's SAIL databank [33], Canada's ICES database [52], USA's eMerge database site [25,53], and the recent Finland's Findata data access regime [1,24]. Some of these warehouses house other forms of data (such as genetic) that is linked to clinical data, however, our discussion focuses on the clinical aspect. The examples aim to illustrate and provide insights into how clinical data access regimes are designed and managed.

## 3.1. ICES

ICES or the Institute for clinical evaluative sciences is a not-for-profit research institute with data covering more than 14 million people in Canada's most populated province, Ontario. Individuals living in Ontario are given a unique health identifier (OHIP) which is recorded in every patient health encounter. Under Ontario's privacy law (PHIPA [54]), ICES is designated as a "prescribed entity", allowing it to receive identifiable health data from health information custodians without patients' consent for the purpose of analysis, assessment or monitoring of health services. In turn, Prescribed entities are required to maintain robust security measures to protect the privacy of the data under the direction of Ontario's Privacy Commissioner.

ICES also maintains other forms of data such as vital statistics, clinical trial data, survey data, transportation data, education data and immigration data. Some of this data does not include the OHIP identifier and are linked locally using deterministic and probabilistic record linkage [55]. ICES has a maintained copy of the registered persons database (RPDB) from the ministry of health. RPDB includes OHIP numbers, demographics and identifying information for all Ontarians. It is used by ICES to aid in the record linkage process.

Only designated and highly qualified ICES employees have access to the original identifiable data (these employees receive intensive training and sign stringent data agreements). They are tasked to link the data, generate a unique ICES identifier for each record, and remove direct identifiers (variables that uniquely identify an individual such as names, phone numbers, IDs,..), thus creating a single de-identified repository. ICES analytic staff have direct access to the de-identified repository for preparing the datasets of approved projects (data requests).

ICES created a public advisory council with members of the public to get feedback on the use of their data for research and ensure that their values and perspectives are taken into consideration. In addition, it has a research ethics board to evaluate data request applications and a Privacy and legal Office to perform a privacy impact assessment on the requests.

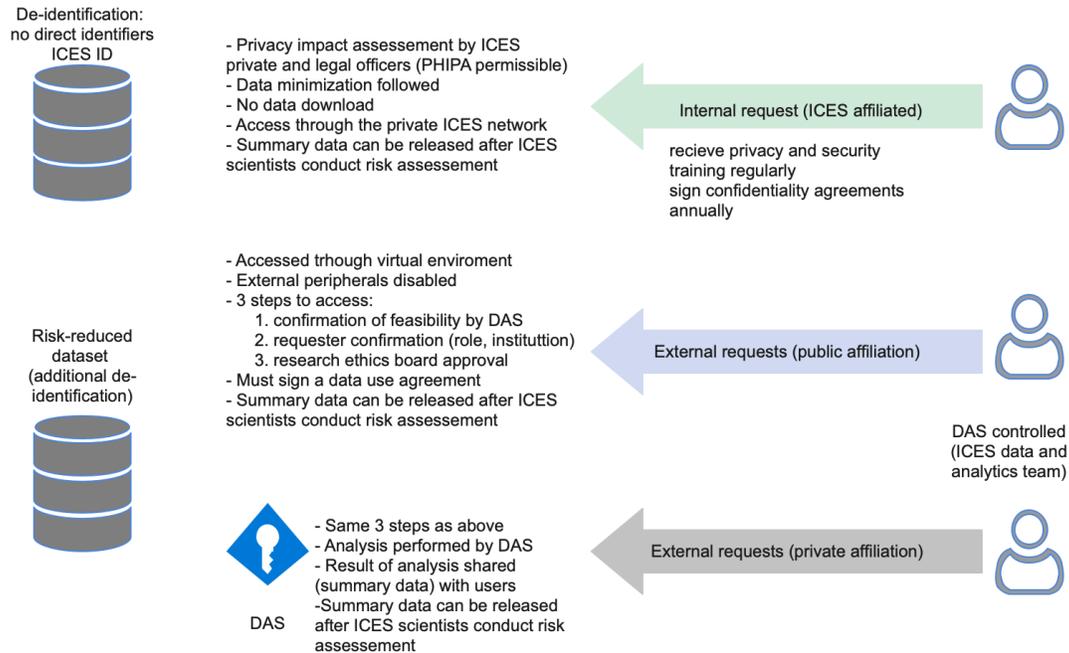

Figure 8. ICES (Institute for clinical evaluative sciences ) research platform overview. DAS refers to ICES data and analytics team.

ICES offers a risk-based data access framework, Figure *8*. Users are distinguished based on their affiliations: external requests originating from a public organization, external requests originating from the private sector and internal requests. External data requests are attended to by an internal team of analysist (ICES data and analytic team, DAS), while internal requests are reviewed by the privacy and legal office.

*ICES-affiliated researchers* may access the data without research ethics board approval, but only after a privacy impact assessment is conducted by ICES privacy and legal office to ensure its adherence with the law (PHIPA), and to minimize risk. For ICES researchers, access is granted to the de-identified data, and is restricted to data that is deemed necessary to achieve the research objectives. *ICES trainees* may also be allowed access to further de-identified data by following the same process.

*Public sector investigators* may access highly de-identified data through a virtual analytics environment. Users submit a project request form to DAS for eligibility and feasibility assessment, if approved, the user needs to seek an additional approval from ICES's research ethics board. Once approvals are secured, the required data is extracted and de-identified further to reduce its risk. External users can access the de-identified data remotely along with analytical software tools and resources. Users may request to download aggregate output data, in which case the data is assessed for privacy risk before being shared with the research team.

*Private sector investigators* may not access individual level data. Instead, they are eligible to access data analysis results performed on their behalf by ICES. In other words, they are eligible for receiving summary data related to their analysis of interest.

Any of the users intending to share results with the public should undergo a risk-assessment to confirm that no personal information can be derived from the published data.

## 3.2. Sail

The secure anonymized information linkage (SAIL) databank is a de-identified dataset about the population of Wales. It was established in 2007 by the population data science group at Swansea University with the objective of improving health and public service delivery through research [33]. Initially, researchers were able to request access to individual level health data. The data scope was later expanded to include other administrative data (such as education, housing and employment) and emergent health data types (such as genomics and imaging) [33].

The data access model is depicted in Figure 9. Data providers separate their data into two parts: a demographic part (name, address, date of birth, gender, and national health number), and remaining data (such as medical data and education data). The demographic data is sent to a trusted third party, the National health service in Wales for informatics service, NWIS, for processing. While the remaining data is sent to SAIL repository located at Swansea University. NWIS maintains a registry of the Welsh population which is used create a pseudonym, referred to as the anonymous linking field, ALF, as well as updating the registry reliably as new data is received (note that ALF is obtained by encrypting the national health service number). The NWIS sends the ALF along with de-identified demographic data containing gender, week of birth, and aggregated residence areas (of population more than 1500) along with a unique key for each residential address (RALF). The purpose of RALF is to be able to identify families/residents of the same house.

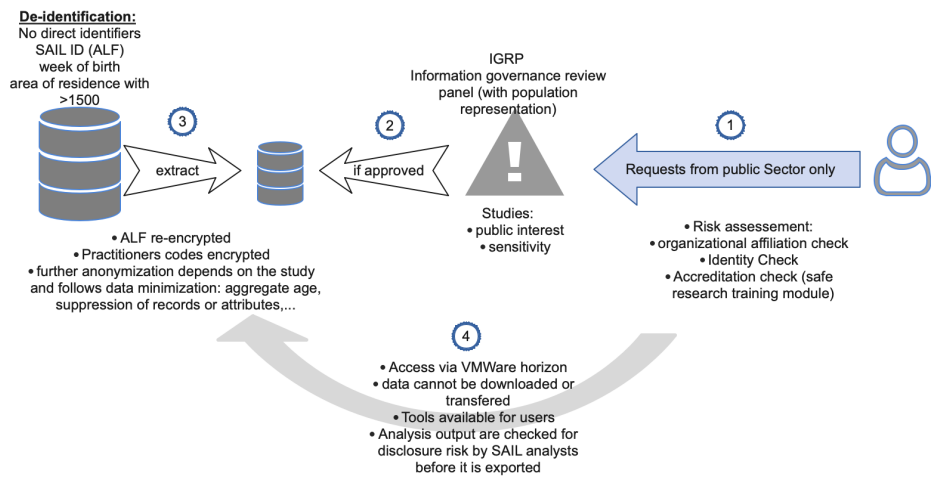

Figure 9. SAIL (or the secure anonymized information linkage ) research platform. ALF refers to the anaonymous linking field.

Data requests are first circulated to an independent information governance review panel (IGRP), which includes representatives from the public as well as professional bodies. The request proceeds only if a favorable decision is reached.

SAIL restricts data access to researchers that are affiliated with the public sector. Private sector researchers are required to collaborate with a public sector organization (including SAIL) through which they can access the data.

Once IGRP approval is obtained, an assessment for the affiliate organization and an identity check is done to ensure credibility. After which the user is allowed access to the data after taking the necessary training and signing the required data sharing agreements. The requested data subset is extracted, ALF is re-encrypted and further de-identification is applied to the data following the principle of data minimization (ages and locations can be further aggregated, and records/fields suppressed).

The data access is done remotely using a remote desktop protocol. while the data remains in the secure platform of SAIL. Data is released in rare cases when accredited organization are involved such as the UK Biobank [56] and subjects' consent is obtained.

As mentioned earlier, SAIL does not handle identifiable data, however, as the data still retains high utility, it is considered potentially identifiable according to GDPR. The framework relied on the GDPR's articles 6 and 9 for a "task carried out in the public interest" in its forming the database. Moreover, as researchers access (further) de-identified data, SAIL is not required to obtain additional (per study) consent from participants.

### 3.3. Vanderbilt

Vanderbilt project is part of the Electronic Medical Records and Genomics (eMERGE) network [32]. eMERGE is a National Institute of Health Initiative and is a consortium of several medical research institutes that combine DNA repositories and EMR systems for advancing biomedical research. Its purpose includes research facilitation, cohort identification, education, and assessment of quality of care.

The Vanderbilt research data warehouse is composed of two data layers: the identified data layer (RD) and the de-identified data layer (synthetic derivative or SD). RD consists of identifiable structured clinical data and is securely stored in the data center under the governance of Vanderbilt IRB. Patients' records have a unique medical records number MRN and are updated daily with new incoming patient data. Unstructured data (such as medical notes and lab reports) are stored separately while still connected through the MRN.

SD is a de-identified research-ready image of RD. The de-identification applied follows the provisions of HIPAA safe-harbor [54]. MRNs are replaced with a random *research unique identifier, RUI* (obtained by one way hashing of the MRN), all patients'/individuals' identifiers are removed in accordance with HIPAA safe-Harbor privacy rule, all service dates are shifted backwards by a random number between 0 and 365 days, that is constant within each patient record (thus preserving time dependence between each patient's medical services), and unstructured data is stripped from all identifiers (Table 2). SD is linked to a de-identified DNA biobank [25]. The facility offers research and informatics support for investigators from data extraction to analytics support with a fee for service.

Researchers are able to access a query functionality through the web. It allows them to generate queries and obtain approximate record counts. It does not require users to obtain pre-approvals and is thus useful for hypothesis generation and study feasibility analysis.

The aforementioned de-identification for the SD reduces the re-identification probability, but does not eliminate it completely. Accordingly, data requests require an approval from the IRB and the signing of a data use agreement (i) that defines the type of analysis allowed on the data, (ii) prohibits re-identification attempts and (iii) details oversight to be applied on the users. After the requested data is extracted, RUIs are substituted with unique identifiers that are investigator based (substitution mechanism is unique for every investigator).

Epidemiological studies that require identifiable or temporal information (such as epidemics) may request access to RD data. Access is granted to the minimal data set required by the study (data minimization) after obtaining an IRB approval and after signing an all-embracing data use agreement.

Table 2. Data de-identification, HIPAA is the health insurance portability and accountability act.

| Identifiers | Action |
| --- | --- |
| 16 HIPAA identifiers: Names, telephone numbers, IDs, email,… | removed |
| MRN | Hashed (pseudonym) |
| Address | Generalized to state |
| Dates of birth, death | Year of birth/death |
| Other dates | Shifted backwards using the same random number per record |

Vanderbilt employs an opt out consent model, where patients are automatically included in the research data unless they opt out. As a counter measure, an inclusive community engagement model was followed during the inception and design of the framework. Then, a community advisory board was created to monitor the ethical conduct of the framework on one hand and to voice community concerns on the other [37]. As RUI are irreversible, it is not possible to inform patients of any actionable findings/breakthroughs resulting from using their samples.

### 3.4. Findata

The Finnish health and data permit authority, Findata, is another recent initiative for facilitating the sharing of health data in Finland. Findata is an independent unit within the Finnish institute for health and welfare and is governed by a dedicated data access law, the *Act on Secondary Use of Health and Social Data,* ASU, which mandates participation of both public and private health care providers in Findata's access regime. The Findata access regime is transparent and comprehensive which makes it an ideal candidate of study.

Finland has a long history of health and social data collection that dates back to 1749 [24]. The ASU, voted into law in May 2019, aims to securely combine health and social care data and to regulate access to this rich data while respecting individuals' privacy rights and expectations.

The Findata platform offers two types of access to the data (Figure 10), *data requests* and *data permits*. Data requests provide access to individual level pseudonymized data, while data permits provide users with aggregated and de-identified statistics.

Users submit their *data utilization plans* online using Findata's *request management system*. The plan should include the research proposal, the purpose of the study (that must fit into 7 pre-defined and specific purposes stipulated by ASU, Table 3), a justification for the specific data requested, the controllers and processors of the data (as per GDPR), an approval from the applicant's institution IRB (if applicable), the legal ground for processing the data along with the security and privacy measures that are followed during the period of data usage.

Findata is required to respond to (complete) data permit applications within 3 months from their initiation, no such deadline is specified for data requests. Once a user's request is accepted, the requested data is gathered from its sources (data holders). Such sources are required to provide the data within 30 days. Once data is received, Findata links the multi-source data using the unique Finnish ID number. Then the data is either pseudonymized and de-identified for data permits, or aggregated and de-identified for data requests.

Users are allowed to download aggregated data obtained through a data request. However, for data permits, the data is made available through a secure environment. Such access is limited to a pre-defined time period, after which all data pertaining to the project are erased from Findata's environment (including links between pseudonyms and Finn ID numbers). Users are required to submit their study results to Findata prior to publication to check whether they are anonymous.

If research results in clinically significant information, i.e. information that may lead to improvement in some participants' quality of life, then Findata may notify them of the finding provided that the participants have consented to receive such information.

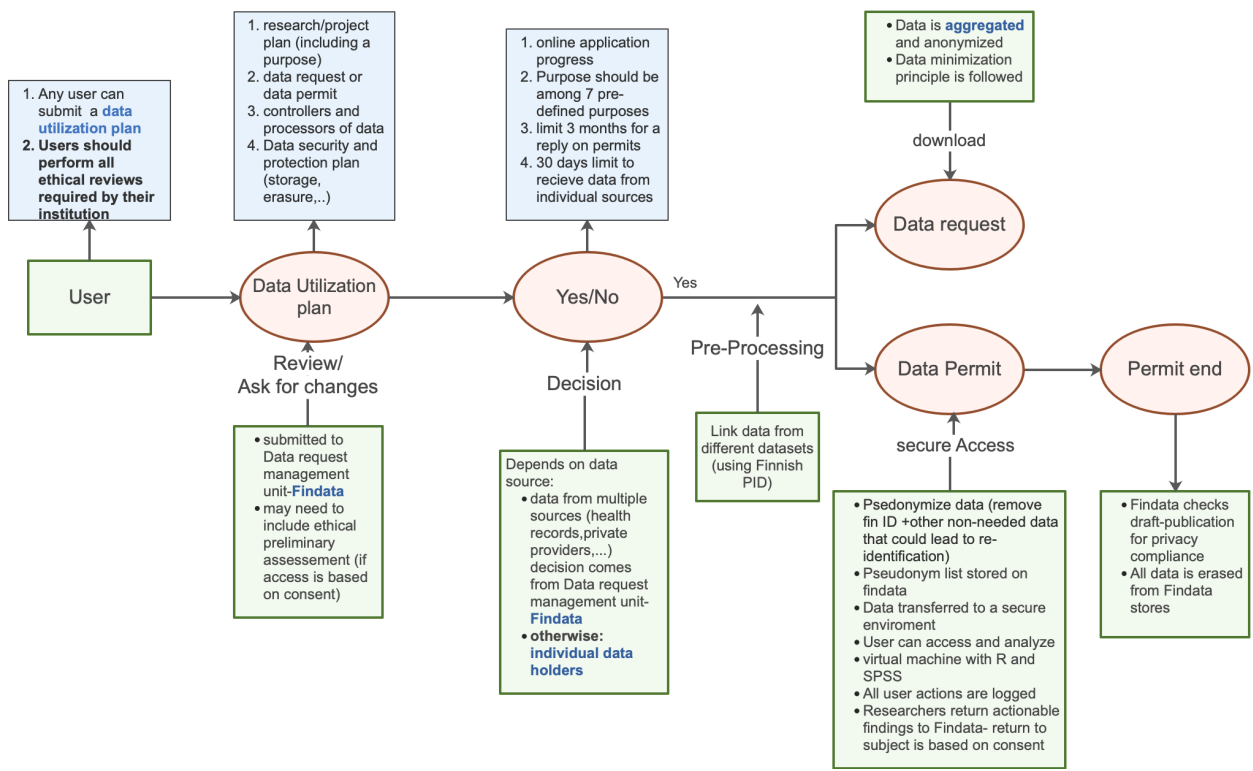

Figure 10. Findata data access platform

Findata is governed/supervised by multiple internal and external parties including the ministry of social affairs and health and the Finnish data protection authority.

In compliance with GDPR, Findata follows the six data protection principles specified in article 5. In particular, Lawfulness of their data processing operations is achieved on the ground of "necessity of performance of a task carried out in the public interest or for the exercises of official authority".

Individual citizens may opt out from use of their data which results in their exclusion from future requests unless overruled by Findata. Findata gets the mandate to reject subjects' opt-out following Article 31 (of the data protection act) only in cases where scientific research is threatened by subjects' opt-out.

Table 3. Purposes specified by ASU (act on secondary use of health and social data) as ground to access data

| Seven ASU approved purposes for data access |
| --- |
| 1. Scientific research |
| 2. Statistics |
| 3. Development and innovation activities |
| 4. Education |
| 5. Knowledge-based management |
| 6. Steering and supervision of social and healthcare by authorities |
| 7. Planning and reporting duties by an authority |

For more information readers are referred to [1,24].

## 4. Discussion

Health data evolved tremendously in the last decades in terms of size, quality, and diversity. It is increasingly collected in electronic health records and updated continuously with every medical visit. It is aggregated with data from other emerging sources including genomics, mobile health devices, and clinical trials. The resulting data is diverse and can serve in unlimited research endeavors. However, access to this data has become increasingly difficult. On one hand data privacy laws are becoming more restrictive and on the other hand, the information collected about individuals in these databases are highly detailed (and unique) making data protection more challenging than ever.

This paper identifies the key challenges in designing a health data access framework and examines how similar issues have been tackled in four existing platforms. Table *4* below summarizes key features in these four use cases.

Table 4. Key features related to the four use-cases, - indicates lack of available information.

|  | ICES | SAIL | Vanderbilt | Findata |
| --- | --- | --- | --- | --- |
| Population | Ontario | Wales | Vanderbilt | Finland |
| Law | PHIPA | UK Data Protection Act [57], GDPR | Common rule [58], HIPAA | GDPR, ASU |
| Data sharing | Permitted | Permitted | Permitted | Mandatory |

| Data type | Health records, clinical trial data, cohorts and registries, statistics, surveys, education, disability, genomics | Health records, education, housing, employment, genomics and imaging | Health records, registries, genetic bio-samples | Health data and social data |
|---|---|---|---|---|
| Governance model | Tiered | Tiered | Tiered | Tiered |
| Consent | Not required | Opt-out | Opt-out | Opt-out (unless overruled by Findata for a specific study) |
| De-identification | Multiple Generalizations | Generalization | Generalization + date obfuscation | No database - data is fetched per request and deleted eventually |
| Access model | 2 de-identified databases (*Risk-based* model- affiliation based) + analysis results for external researchers | One de-identified database + additional de-identification based on study requirements (*mixed model*) | One de-identified database (*One size fits all*) + DUA (Possible access to identifiable data if required by study). | No maintained database *Adaptive model* (on the basis of data-minimization) |
| Research programs priority | Yes | yes | yes | yes |
| Reversible pseudonym | - | yes | no | Yes |
| Users | *No restriction* on users (affiliation impacts utility of data received) | *Restricted* to researchers affiliated with public institutions | *No restriction* | *No restriction* |
| Access mode | *Virtual access to record-level data* (study results can be downloaded after risk-assessment) *Download for Summary data* | *Virtual access to record-level data* (study results can be downloaded after risk-assessment) | *Data downloads* (record level data)+data use agreements | *Virtual access for record-level data* (study results can be downloaded after risk-assessment) *Download for aggregate data* |
| Exploratory analysis | - | Requires approval | No approval required | Requires approval |

The results indicate a decreasing reliance on informed consent (consent prior to each data usage) in favor of *broader consent mechanisms* such as opt out. At the same time, in response to the reduction in consent precision, community involvement in the early stages of framework design are becoming more prevalent in addition to the involvement of community members in oversight and ethical boards. A good way forward would be to design legislations to protect participants against discrimination, should they be subjected to data leaks.

Additionally, and in compliance with ethical data sharing principles [59], participants should be given the possibility to manage and control their clinical data, should they wish to take this role. In such cases, participant must acknowledge the weight of such task – as effectively managing their data requires informed decisions on complex issues that may require considerable time to be fully comprehended and assessed [27].

Another trend in modern frameworks is the shift from data download to *virtual, and time-bound, data access* supplemented with *critical safeguards and stronger (multi-layer) oversight mechanisms.* This is justified by the need to maintain control and governance over the data at all times. A more inclusive way forward would be to come up with a trust rating system for research institutions/users. The rating would assist data holders in objectively assessing the risk posed by a data requests.

However, a trust system should not rescind the need for *stringent data sharing agreements*. In a recent example (2016), NHS partnered with DeepMind to utilize their AI capabilities for acute kidney injury management [60]. Subsequently, Google acquired DeepMind's project and inherited the partnership, effectively transferring control over the patients' data to the United States [61] and obliterating participants' trust in other projects.

22. Mallon AM, Häring DA, Dahlke F, Aarden P, Afyouni S, Delbarre D, et al. Advancing data science in drug development through an innovative computational framework for data sharing and statistical analysis. BMC Med Res Methodol. 2021 Nov 14;21(1):250.

23. Kaye J, Hawkins N. Data sharing policy design for consortia: challenges for sustainability. Genome Med. 2014 Jan 29;6(1):4.

24. Ausloos J, Leerssen P, ten Thije P. Operationalizing Research Access in Platform Governance: What to Learn from Other Industries? 2020 Jun 25 [cited 2022 Nov 4]; Available from: https://dare.uva.nl/search?identifier=90e4fa77-d59a-49f1-8ccd-57d0725122bd

25. Danciu I, Cowan JD, Basford M, Wang X, Saip A, Osgood S, et al. Secondary use of clinical data: The Vanderbilt approach. J Biomed Inform. 2014 Dec 1;52:28–35.

26. Health Data Governance: Privacy, Monitoring and Research | en | OECD [Internet]. [cited 2023 Feb 15]. Available from: https://www.oecd.org/publications/health-data-governance-9789264244566-en.htm

27. Dankar FK, Gergely M, Malin B, Badji R, Dankar SK, Shuaib K. Dynamic-informed consent: A potential solution for ethical dilemmas in population sequencing initiatives. Comput Struct Biotechnol J. 2020 Jan 1;18:913–21.

28. El Emam K, Rodgers S, Malin B. Anonymising and sharing individual patient data. bmj. 2015;350:h1139.

29. Garfinkel S. De-Identification of Personal Information [Internet]. National Institute of Standards and Technology; 2015 Oct [cited 2022 Dec 8]. Report No.: NIST Internal or Interagency Report (NISTIR) 8053. Available from: https://csrc.nist.gov/publications/detail/nistir/8053/final

30. Dankar FK, Al-Ali R. Building of a Large Scale De-Identified Biomedical Database in Qatar-Principles and Challenges. Qatar Found Annu Res Conf Proc. 2016 Mar 1;2016(1):HBPP3324.

31. Pulley J, Clayton E, Bernard GR, Roden DM, Masys DR. Principles of Human Subjects Protections Applied in an Opt-Out, De-identified Biobank. Clin Transl Sci. 2010;3(1):42–8.

32. All About The Human Genome Project (HGP) [Internet]. National Human Genome Research Institute (NHGRI). [cited 2017 Feb 26]. Available from: https://www.genome.gov/10001772/All-About-The--Human-Genome-Project-HGP

33. Jones KH, Ford DV, Thompson S, Lyons RA. A Profile of the SAIL Databank on the UK Secure Research Platform. Int J Popul Data Sci. 2019 Nov 20;4(2):1134.

34. Vatsalan D, Christen P, O'Keefe CM, Verykios VS. An evaluation framework for privacy-preserving record linkage. J Priv Confidentiality. 2014;6(1):3.

35. El Emam K, Jonker E, Fineberg A. The Case for De-identifying Personal Health Information. Soc Sci Res Netw [Internet]. 2011 [cited 2012 Sep 18]; Available from: http://papers.ssrn.com/sol3/papers.cfm?abstract_id=1744038